\documentclass[epj]{svjour}
\usepackage{graphics}
\usepackage{multirow}
\begin{document}
\title{A Novel Method for the Measurement of Half-Lives and Decay Branching Ratios of Exotic Nuclei}
\author{Ivan Miskun\inst{1}\thanks{Part of doctoral thesis, Justus Liebig University Gie\ss en, in preparation.} \and Timo Dickel\inst{1,2} \and Israel Mardor\inst{3,4} \and Christine Hornung\inst{1} \and Daler Amanbayev\inst{1} \and Samuel Ayet San Andr\'{e}s\inst{1,2} \and Julian Bergmann\inst{1} \and Jens Ebert\inst{1} \and Hans Geissel\inst{1,2} \and Magdalena G\'{o}rska\inst{2} \and Florian Greiner\inst{1} \and Emma Haettner\inst{2}  \and Wolfgang R. Pla\ss\inst{1,2} \and Sivaji Purushothaman\inst{2} \and Christoph~Scheidenberger\inst{1,2} \and Ann-Kathrin~Rink\inst{1} \and Helmut Weick\inst{2} \and Soumya Bagchi\inst{1,2,6} \and Paul Constantin\inst{5} \and Satbir Kaur\inst{6} \and Wayne Lippert\inst{1} \and Bo Mei\inst{5} \and Iain Moore\inst{7} \and Jan-Hendrick Otto\inst{1} \and Stephane Pietri\inst{2} \and Ilkka Pohjalainen\inst{7} \and Andrej Prochazka\inst{2} \and Christoph Rappold\inst{1,2} \and Moritz P.~Reiter\inst{1,8} \and Yoshiki K. Tanaka\inst{2} \and John S. Winfield\inst{2} \and for the Super-FRS Experiment Collaboration} 

%
\offprints{Israel Mardor, mardor@tauex.tau.ac.il}          
\institute{II.~Physikalisches Institut, Justus-Liebig-Universit\"at Gie\ss en, 35392 Gie\ss en, Germany \and GSI Helmholtzzentrum f\"ur Schwerionenforschung GmbH, 64291 Darmstadt, Germany \and Tel Aviv University, 6997801 Tel Aviv, Israel \and Soreq Nuclear Research Center, 81800 Yavne, Israel \and IFIN-HH/ELI-NP, 077126, M$\breve{a}$gurele - Bucharest, Romania \and Saint Mary's University, NS B3H 3C3 Halifax, Canada \and University of Jyv\"askyl\"a, 40014 Jyv\"askyl\"a, Finland \and TRIUMF, BC V6T 2A3 Vancouver, Canada}
\date{Received: date / Revised version: date}
%

\abstract{
A novel method for simultaneous measurement of masses, Q-values, isomer excitation energies, half-lives and decay branching ratios of exotic nuclei has been demonstrated. The method includes first use of a stopping cell as an ion trap, combining containment of precursors and decay-recoils for variable durations in a cryogenic stopping cell (CSC), and afterwards the identification and counting of them by a multiple-reflection time-of-flight mass spectrometer (MR-TOF-MS). Feasibility has been established by recording the decay and growth of $^{216}$Po and $^{212}$Pb (alpha decay) and of $^{119m2}$Sb ($t_{1/2}=850\pm90$ ms) and $^{119g}$Sb (isomer transition), obtaining half-lives and branching ratios consistent with literature values. Hardly any non-nuclear-decay losses have been observed in the CSC for up to $\sim$10 seconds, which exhibits its extraordinary cleanliness. For $^{119}$Sb, this is the first direct measurement of the ground and second isomeric state masses, resolving the discrepancies in previous excitation energy data. These results pave the way for the measurement of branching ratios of exotic nuclei with multiple decay channels. 
\PACS{
      {21.10.Dr }{Binding energies and masses} \and
      {21.10.−k}{Properties of nuclei; nuclear energy levels} \and
      {21.10.Tg}{Lifetimes, widths} \and
      {23.90.+w}{Other topics in radioactive decay and in-beam spectroscopy} \and
      {27.60.+j}{90 ≤ A ≤ 149} \and
      {29.30.−h}{Spectrometers and spectroscopic techniques} \and
      {29.90.+r}{Other topics in elementary-particle and nuclear physics experimental methods and instrumentation}
     } 
} 

\authorrunning{I. Miskun et al.}
\titlerunning{A Novel Method for Measuring Half-Lives and Decay Branching Ratios...}
\maketitle
\section{Introduction}
\label{intro}
Half-lives and branching ratios to various decay channels are important properties of unstable exotic nuclei, as they provide vital input to the verification and extension of nuclear structure models throughout the nuclear chart, and to astrophysical nucleosynthesis models \cite{Iliadis2015}.  
Exotic nuclei, especially far from the stability valley, likely have numerous decay channels, which may include $\alpha$ and $\beta$ decays, spontaneous fission, direct single- or multi-nucleon emission, and $\beta$-delayed fission and nucleon emission. For a review on $\beta$-delayed decays and their importance to nuclear physics and astrophysics, see Ref. \cite{Borge2013}.

Half-lives and decay branching ratios are mainly measured by the detection of the relevant light particles involved in the decay ($\alpha$, $\beta$, nucleons) as a function of time, and the identification of the recoil nuclei via their $\gamma$ lines, as they are usually created in excited states \cite{Dillman2018,Lorusso2012,Mukha2006}.
Penning traps are powerful for purification of precursor beams, whose decay can be researched via several post-trap spectroscopy techniques with minimal background \cite{eronen2016}. 

Other methods use in-trap nuclear decay, and include direct mass measurements of precursors and/or recoils via a Penning trap \cite{Herlert2012} or a multi-reflection time-of-flight mass-spectrometer (MR-TOF MS) \cite{Wolf2016}.
In another approach, different decay branches are identified and counted by measuring the momentum of the recoil nucleus via time of flight, after decay inside a Paul trap \cite{Yee2013}.
It has further been proposed to measure decay branching ratios with storage rings \cite{Evdokimov2012} and by a series of gas-filled stopping cells and MR-TOF-MSs \cite{Miyatake2018}.

In this paper we describe a novel method for measuring half-lives and decay branching ratios of exotic nuclei. We demonstrate for the first time use of a stopping cell as an ion trap. Containment times up to $\sim$10 seconds are shown, indicating the high cleanliness level of this cell. The method is universal, and can be used for nuclides that decay via one or more of any of the channels that are listed above.

We require only very general prior knowledge of nuclear properties, since all are measured during the experiment. This is accentuated in the current article, where while demonstrating our method, we measured directly for the first time the mass of the ground and the second isomeric state ($^{119m2}$Sb, $t_{1/2}=850\pm90$ ms) of $^{119}$Sb, offering a solution to a long-existing conflict regarding its excitation energy and spin assignment \cite{Shroy1979,Lunardi1987,Porquet2005,Moon2011}.

The paper is organised as follows: in Section~\ref{sec:2} we describe our novel method for measuring half-lives and branching ratios in the FRS-IC; we present the results of our feasibility study in Section~\ref{sec:3}; Section~\ref{sec:4} discusses ongoing developments of the method, and lists future possible measurements with it. We conclude in Section~\ref{sec:5}.

\section{Outline of methodology}
\label{sec:2}
Our method is implemented in the Fragment Separator (FRS) Ion Catcher (IC) \cite{Plass2013} at the FRS \cite{Geissel1992b} , which comprises a gas-filled cryogenic stopping cell (CSC) \cite{Ranjan2015}, an RFQ-based mass filter and beam line \cite{Plass2007}, and an MR-TOF-MS \cite{Dickel2015b}. 

\begin{figure*}[ht]
\resizebox{1\textwidth}{!}{%
  \includegraphics{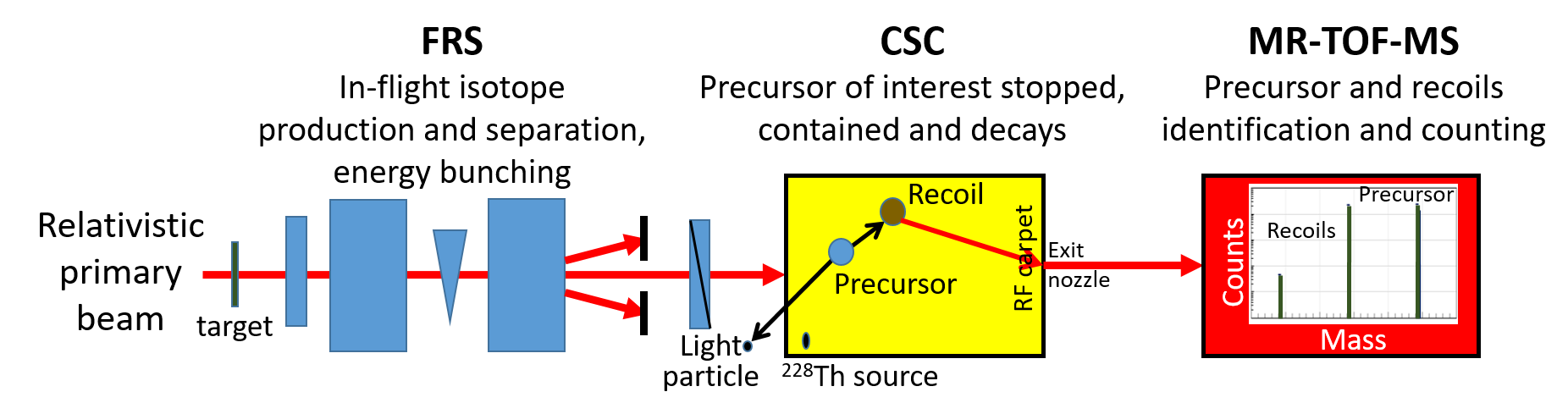}
}
\caption{Scheme of the measurement method. A single decay event inside the CSC and the accumulated mass spectrum in the MR-TOF-MS after the extraction of precursors and recoils are depicted. The internal $^{228}$Th $\alpha$-recoil source inside the CSC, used for the offline measurements in this work, is also shown. }
\label{fig:0}       
\end{figure*}

The general concept is outlined in Fig.~\ref{fig:0}. A primary beam is accelerated to relativistic energy by the GSI SIS-18 synchrotron. A pure ensemble of precursors is produced and separated in flight via the FRS. The precursors are contained in the CSC for a controllable duration, while they decay. Due to the high CSC gas density, almost all recoil isotopes are contained. The precursor and recoil isotopes are consequently extracted to the MR-TOF-MS, where they are identified by their mass-to-charge ratio and counted. The precursor decay branching-ratios are determined from the ratios between the number of recoils. 

In addition to branching ratios, we simultaneously measure precursor and recoil masses (and thus the precursor Q-values), and the precursor half-life. To reduce systematic errors, half-lives and branching ratios are studied as a function of CSC containment time. Such a method is uniquely-possible with the Ion Catcher at the FRS, because the high energy of the FRS fragments enables the stopping of a clean sample of the ions of interest in the CSC.

\subsection{Obtaining clean precursor samples in the CSC}
\label{sec:2.1}
\begin{figure}[ht]
\resizebox{0.48\textwidth}{!}{%
  \includegraphics{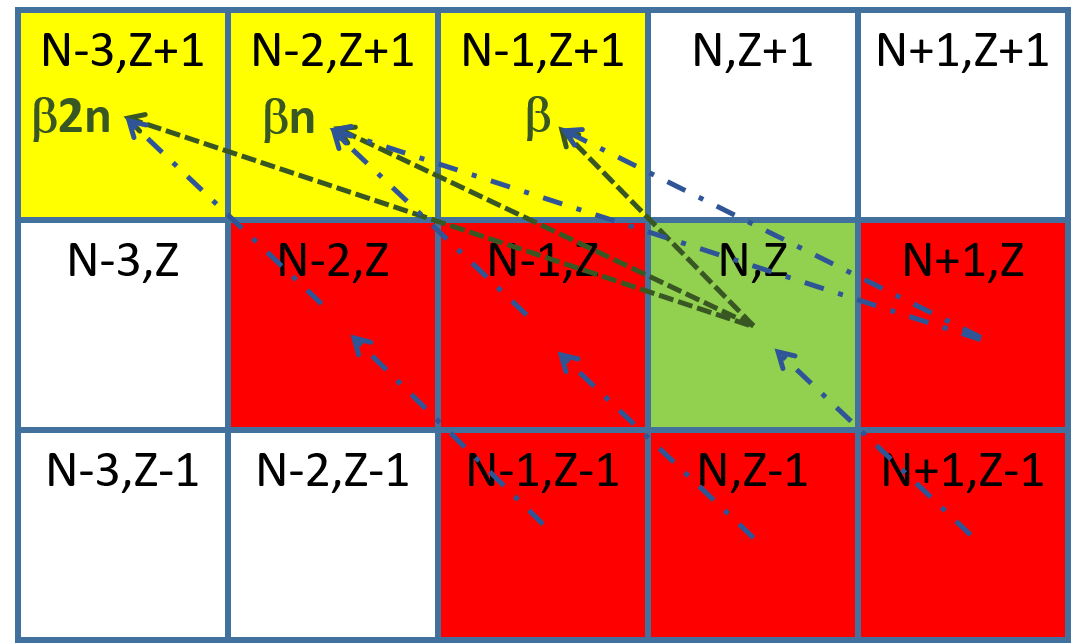}
}
\caption{Visualization of the masking effect of contamination nuclides on the branching ratio measurement in the case of $\beta$-delayed neutron emission. The green nuclide is the precursor of interest, and the yellow ones are the recoils of interest from the possible branches ($\beta$, $\beta$n, $\beta$2n), indicated by dashed arrows. Beam contamination of these recoils will obviously interfere with the measurement. In addition, contamination of the red nuclides, which might decay by $\beta$, $\beta$n and $\beta$2n to the recoils of interest (either directly or in a chain), indicated by dot-dashed arrows, will mask those that result from the precursor of interest. Contamination by white nuclides will not affect the results and can thus be tolerated.}
\label{fig:01}       
\end{figure}

A major requirement of our method is to have a pure sample of precursor ions stopped in the CSC. Our main concern are neighboring nuclides that are either the eventual recoil isotopes of the decay of interest, or nuclides that decay into the precursor or its recoil isotopes. Such effects might mask the investigated precursor and recoils at a level that will render the method inadequate. A visualization of masking for the case of $\beta$-delayed neutron emission is given in Fig.~\ref{fig:01}, indicating which are the nuclides that should not be stopped in the CSC directly from the fragment beam. Similar identification of unwanted contaminant nuclides can be deduced for any other nuclear decay process.

%

The FRS can provide an isotope beam with the required cleanliness due to the high energy of its fragments. This can be achieved by either using an achromatic mode and closing the slits \cite{Geissel1992b}, or using a monochromatic mode and taking advantage of the stopping thickness in the CSC (a few mg/cm$^{2}$), which will stop only the precursor of interest. LISE++ \cite{Tarasov2008} and MOCADI \cite{Iwasa1997} simulations that were performed for several cases indicate maximal masking of investigated precursors and recoils at a level of $10^{-4}$, setting this value as a lower limit on measurable branching ratios with this method.

Another potential source for contaminants is generated by neutron removal reactions of the incoming beam upstream of the CSC. Such reactions generate neutron-deficient isotopes of the investigated precursor. However, the majority of these contamination isotopes will not be stopped in the CSC. The upper limit of this contribution is estimated to be $10^{-4}$, for neutron-rich nuclides. 

We thus conclude that for $\beta$-delayed neutron emission, branching ratios as low as $10^{-4}$ are detectable by this method. This lower limit will vary for different decay channels, depending on the relative positions of the precursor and recoils in the nuclide chart.

\subsection{Decay recoils in the CSC}
\label{sec:2.4}
The ranges of recoil nuclei in relevant decay processes in He buffer gas was studied with the stopping and range of ions in matter Monte Carlo program, SRIM \cite{Ziegler2010}, using a typical density of 30~$\mu$g/cm$^3$ (namely 3.17 mg/cm$^2$ for the 105 cm long CSC, which was the areal density for the measurements with beam-generated isotopes in this work). For $\gamma$ decay, the recoil kinetic energy is a few tens of eV, corresponding to ranges of a few $10^{-2}$ mm. In $\beta$ decay, the recoil kinetic energy is a few hundred eV, resulting in a range of 0.05–-0.20 mm. When there is also emission of $\beta$-delayed nucleons (or for direct nucleon emission), the recoil kinetic energy rises to tens of keV, increasing the ranges to 0.5–-2.0 mm. In $\alpha$ decay, the recoil energy is in the 100--200 keV range, leading to ranges of about 10-12 mm for typical $\alpha$ recoil isotopes.

The CSC inner dimensions are 105 cm in length and 25 cm in diameter, so as long as the isotopes of interest are contained and decay in the bulk, there is no risk to lose recoils on the inner walls. However, in the experiments that are described in this paper, containment of the isotopes took place on the CSC RF carpet close to the exit nozzle (see Fig.~\ref{fig:0}). As described in detail in \cite{Ranjan2015}, the pitch of the RF carpet rings is 0.25 mm, and the isotopes are contained in the buffer gas at about half that distance above the RF carpet. 

For $\gamma$-decay there is no effect, since the recoil range is much below $\sim$0.1 mm. On the other hand, for $\alpha$ decay all recoils in the hemisphere towards the RF carpet will be lost, so one merely needs to double the number of recoils counted in the MR-TOF-MS to accommodate for this. These are the decay types that were investigated within the current work.

However, for $\beta$ decay and $\beta$-delayed and direct nucleon emission, the recoil range is in the same order of magnitude as the distance from the RF carpet, rendering a systematic effect that is very difficult to estimate. In Section~\ref{sec:4.1} we describe a solution to this problem, by containment of the isotopes of interest in the bulk for the vast majority of the containment time.

\subsection{Mass measurements and isotope counting}
\label{sec:2.2}
Mass measurements and isotope counting at the FRS-IC are performed with an MR-TOF-MS, which entails a unique combination of performance parameters - fast ($\sim$ms), accurate (relative mass accuracy of $<10^{-6}$) and non-scanning \cite{Dickel2015b}.

The development and validation of the MR-TOF-MS data-analysis procedure is presented in a separate publication \cite{Ayet2019}. Mass values and amounts of the measured isotopes and their uncertainties are obtained after fitting the mass peaks to a Hyper-EMG function \cite{Purushothaman2017}. The procedure allows accurate mass determination even for the most challenging conditions, including very low numbers of events and overlapping mass distributions. In the mass measurements reported in Ref. \cite{Ayet2019}, a total relative uncertainty down to $6\cdot 10^{-8}$ was achieved.
. 

\subsection{Half-life and branching ratio measurement}
\label{sec:2.3}
Half-lives of exotic isotopes are measured directly in the FRS-IC by varying the isotopes' containment time in the CSC. 
The containment time is varied from its minimal extraction time, tens of milliseconds \cite{Plass2013} up to $\sim$10 seconds (see Section 3.2), setting the half-life range that can be measured with this method. This is suitable for a very wide range of exotic isotopes and isomeric states.

Decay branching ratios are derived by calculating the ratios between the amount of the recoil isotope resulting from a specific decay channel and the total amount of recoil isotopes from all decay channels. 

For the measurements with beam-generated isotopes described herein, the spill structure of the heavy-ion synchrotron SIS-18 is adjusted to allow for short (few millisecond) bunches, separated by long ($\sim$10 second) breaks. This separation is chosen to match the longest anticipated storing time inside the CSC. As a result, short spills of ions are injected into the CSC. 

Ion storage in the CSC is achieved by increasing the potential on the extraction nozzle to block extraction for an adjustable duration, and lowering it for a few milliseconds to enable extraction. The containment time is defined as the time between the temporal mid-point of the beam-spill and the start of the low-potential pulse at the extraction nozzle.

For isotopes generated by internal radioactive recoil sources, 'spills' are created by pulsing the potential of the source. It is normally negative with respect to the surrounding electrodes, thereby preventing ions from leaving the vicinity ($\sim$1 mm) of the source, and is pulsed periodically to positive values for a few milliseconds.

After extraction from the CSC, the isotopes are identified and counted by analyzing the MR-TOF-MS spectra via the procedure from \cite{Ayet2019}. The precursor's half-life is extracted by fitting the precursor ($P(t)$) and recoil ($R_i(t)$) counts to the well-known solutions of the Bateman equations \cite{bateman1910}:

\begin{equation} \label{eq:counts_precursor}
P(t) = A_p \cdot e^{- \lambda_P \cdot t}
\end{equation}

\begin{equation} \label{eq:counts_recoil_decay}
R_i(t) = y_{0i}\cdot e^{-\lambda_{Ri}\cdot t} + \frac{\lambda_P}{\lambda_{Ri}-\lambda_P} \cdot n_{ri} \cdot A_p \cdot \left (e^{-\lambda_P \cdot t} - e^{-\lambda_{Ri} \cdot t} \right) \end{equation}
where $A_p$ is the initial amount of the precursor, $\lambda_P$ is the precursor's decay constant, related to its half-life by $\lambda_P=\frac{ln(2)}{t^{P}_{1/2}}$, $\lambda_{Ri} = \frac{ln(2)}{t^{Ri}_{1/2}}$ is the i'th recoil decay constant, $y_{0i}$ is a constant contamination of the i'th recoil isotope, and $n_{ri}$ is the i'th decay branching ratio. $n_{ri}\cdot A_p$ is the amount of the i'th recoil isotope at infinite time (practically, a time much longer than the precursor half-life). The sum of $n_{ri}$ is equal to unity.  

In the current work the half-lives of the recoils are much longer than the precursor's, as well as than the containment times, so for our analysis we assume$\lambda_R \rightarrow 0$ and use the simplified expression: 
\begin{equation} \label{eq:counts_recoil}
R_i(t) = y_{0i} + n_{ri} \cdot A_p \cdot \left (1 - e^{-\lambda_P \cdot t} \right)       
\end{equation}

Note that in the investigation of $\alpha$, $\beta$ and direct and $\beta$-delayed nucleon emission, the FRS will provide an essentially contaminant-free ion sample (Section~\ref{sec:2.1}). However, in the case of $\gamma$-decay of isomeric states, there will always be a non-zero constant contamination, since there is no way to separate isomeric states from ground states in the FRS (Section~\ref{sec:2.1}). The amount of constant contamination ($y_{0i}$) can be obtained directly by a measurement of all recoil candidates with a containment time much shorter than the precursor half-life.

The precursor half-life and decay branching ratios are obtained via a global fit of the precursor and all recoil candidates, thereby providing an improved uncertainty, both statistically and systematically. In particular, the statistics in such a measurement remain constant as a function of containment time, whereas when measuring only the precursor, the statistics decrease as less and less precursors are detected.

By 'global' we refer to a unified fit of the data sets of both the precursor and all recoil candidates, using Eq. ~\ref{eq:counts_precursor} for the precursor data and Eq. ~\ref{eq:counts_recoil_decay} or Eq. \ref{eq:counts_recoil} for the recoil candidates data. We apply the following constraints to the global fit: the same decay constant parameter ($\lambda_P$) for the precursor decay and recoil growth, all branching ratios ($n_{ri}$), constant contaminations ($y_{0i}$) and the precursor initial amount ($A_p$) are non-negative, and the sum of $A_p$ and all $y_{0i}$ is unity, as well as the sum of all $n_{ri}$.

The precursor's half-life and decay branching ratios can be also extracted from a measurement at only one CSC containment time, provided that the recoil constant contaminations are negligible or are known (and can be subtracted from the counts), and the recoils half-lives are relatively long. 
The precursor half-life is obtained by division of Eq.~\ref{eq:counts_precursor} by Eq.~\ref{eq:counts_recoil} (setting $y_{0i}=0$) and subsequent algebraic manipulation: 

\begin{equation} \label{eq:halflife_one_tc}
t_{1/2}=\ln{2} \cdot t_c \cdot \left [\ln{\frac{1+D}{D}} \right]^{-1}
\end{equation}
where $D\equiv \frac{P(t_c)}{\Sigma R_i(t_c)}$, and $t_c$ is the selected CSC containment time.
The branching ratio of the i'th recoil is obtained by

\begin{equation} \label{eq:br_one_tc}
n_{ri}=\frac{R_i(t_c)}{\Sigma R_i(t_c)}
\end{equation}

The statistical uncertainty in the measured half-lives and branching ratios depend on the statistical uncertainty of the isotope counts for each containment time, which are evaluated by the analysis procedure \cite{Ayet2019}.

A possible source for systematic uncertainties might be loss of particles due to chemical reactions (e.g., charge exchange, molecular formation) in the CSC buffer gas. The results in this paper (Section~\ref{sec:3}) indicate that this is a minute effect.

In particular, branching ratio results might be affected systematically by element dependence of these chemical reactions. However, it has been established in previous FRS-IC experiments that the CSC survival and extraction efficiency is essentially element independent, including comparison of a noble element (Rn), and one of the most reactive ones (Th) \cite{Plass2019}.
In case of doubt or for newly-investigated elements, we will measure the temporal behavior of stable or near-stable isotopes of the element under investigation.

\section{Results and discussion}
\label{sec:3}
In the following sub-Sections we present results from both off-line ($\alpha$-recoil source) and on-line (fragment beam) experiments. In all these measurements, the DC pushing field in the CSC was 17 V/cm and the RF carpet voltage was 97 V$_{p-p}$ at a frequency of 5.92 MHz.

\begin{figure}[ht]
\resizebox{0.48\textwidth}{!}{%
  \includegraphics{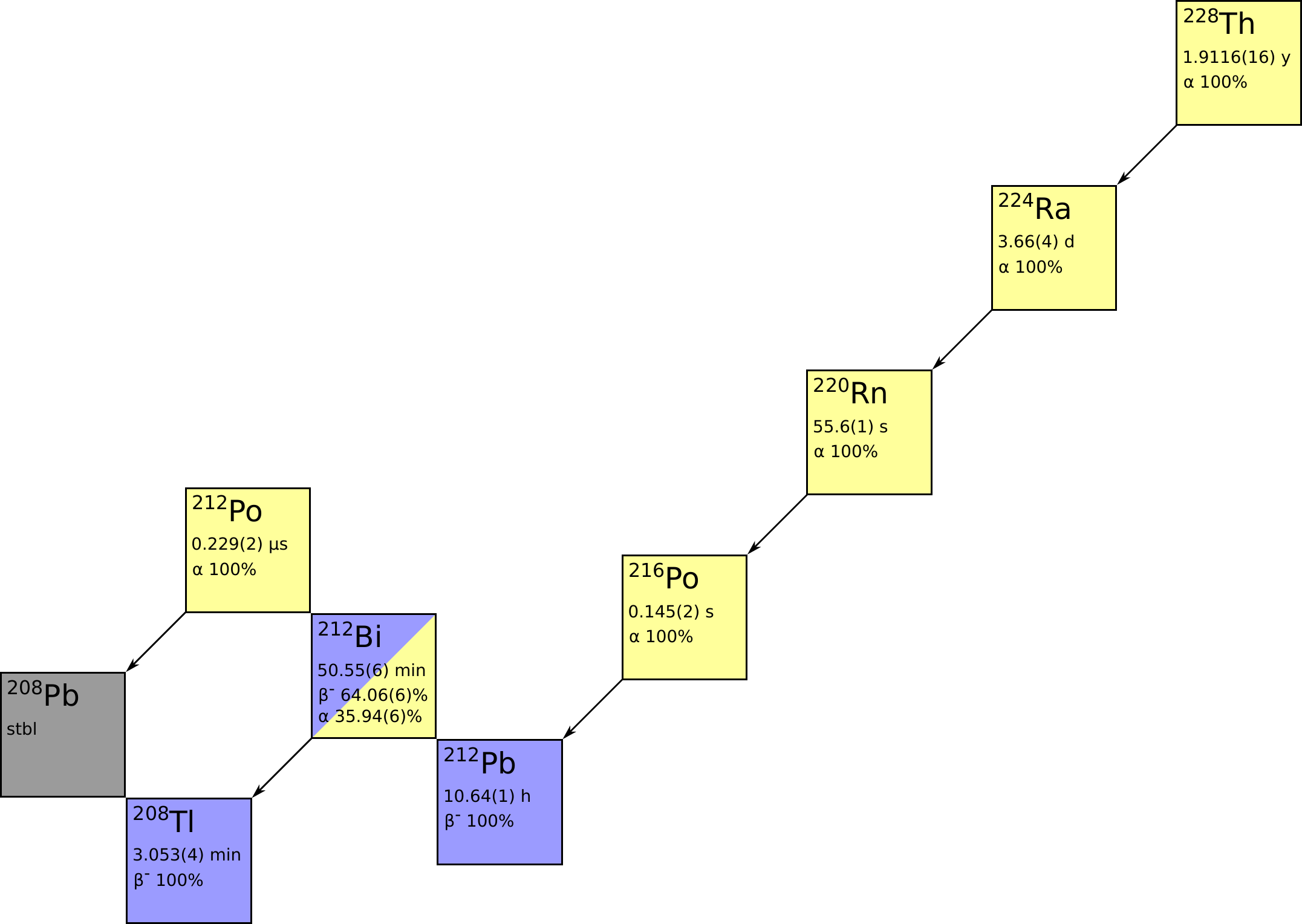}
}
\caption{Decay scheme of $^{228}$Th. Yellow: $\alpha$ emitters; blue : $\beta$ emitters; and gray: stable isotopes.}
\label{fig:228Th-scheme}       
\end{figure}

Offline measurements were performed with the $^{228}$Th $\alpha$-recoil source that is installed inside the CSC (Fig.~\ref{fig:0}). During these measurements, the He gas pressure in the CSC was 33 mbar and its temperature was 94 K, corresponding to an areal density of 1.8~mg/cm$^2$. The decay scheme of this source is shown in Fig.~\ref{fig:228Th-scheme}.

Online measurements were performed with a 300~MeV/u $^{238}$U primary beam from the heavy-ion synchrotron SIS-18 with an intensity of up to $2.5\cdot10^8$ ions per spill. A typical spill length of 200~ms was used, which impinged on a beryllium target with an areal density of 0.270~g/cm$^2$ to generate fragments. Due to the low primary beam energy the matter in the FRS beam-line was minimized. The monochromatic degrader at the mid-focal plane of the FRS had an areal density of 737.1~mg/cm$^2$. During these measurements, the He pressure in the CSC was 64.3 mbar and the temperature was 103 K, corresponding to an areal density of 3.17~mg/cm$^2$. 
A consequence of the low primary beam energy and intensity was that fragment separation was not optimal, generating systematic effects on our data, and signal statistics were generally low, increasing statistical uncertainties.
\subsection{Mass spectra}
\label{sec:3.1}
\begin{figure}[ht]
\resizebox{0.48\textwidth}{!}{%
  \includegraphics{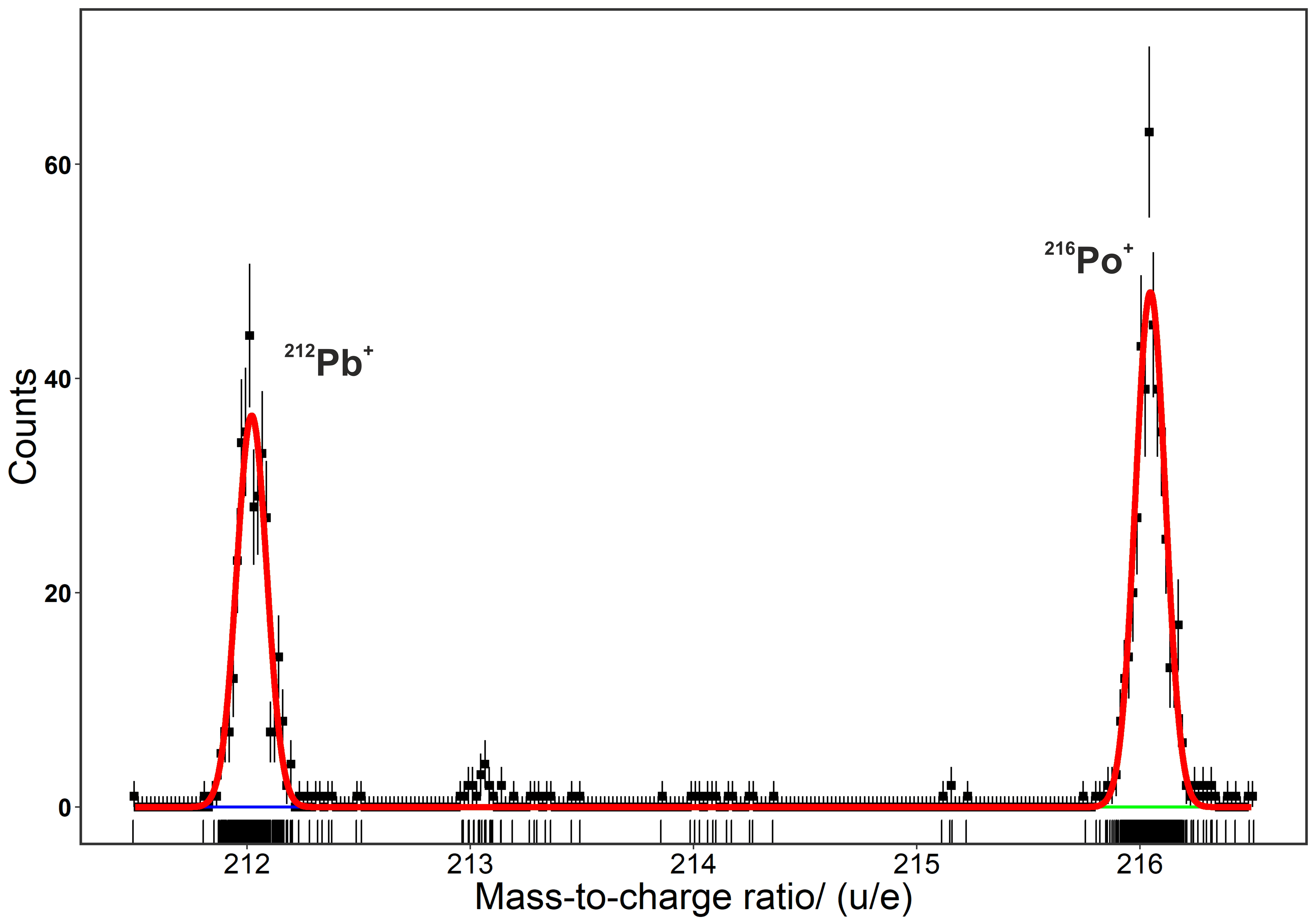}
}
\caption{Mass spectrum from the $^{228}$Th source installed inside the CSC. $^{216}$Po and $^{212}$Pb are identified by their mass. At the other mass numbers a small amount of molecular contaminants can be seen, which are formed due to the high ionization density in the vicinity of the $^{228}$Th source. 
The red line represents the overall fit to the $^{216}$Po and $^{212}$Pb peaks. The fit does not take into account the molecular contaminants between them. The un-binned data is shown as well.}
\label{fig:1}       
\end{figure}

In Fig.~\ref{fig:1} we show a measured mass spectrum of $^{216}$Po and $^{212}$Pb ions, generated by the $^{228}$Th recoil ion source that is installed inside the CSC. The fit curves in Fig.~\ref{fig:1} are the result of the procedure described in \cite{Ayet2019}. Similar spectra and their analysis were used at various containment times to measure directly the half-life of $^{216}$Po and to exhibit the simultaneous decay and growth of the precursor and recoil isotopes in the $\alpha$ decay of $^{216}$Po (Section~\ref{sec:3.3}).
%

\begin{figure}
\resizebox{0.48\textwidth}{!}{%
  \includegraphics{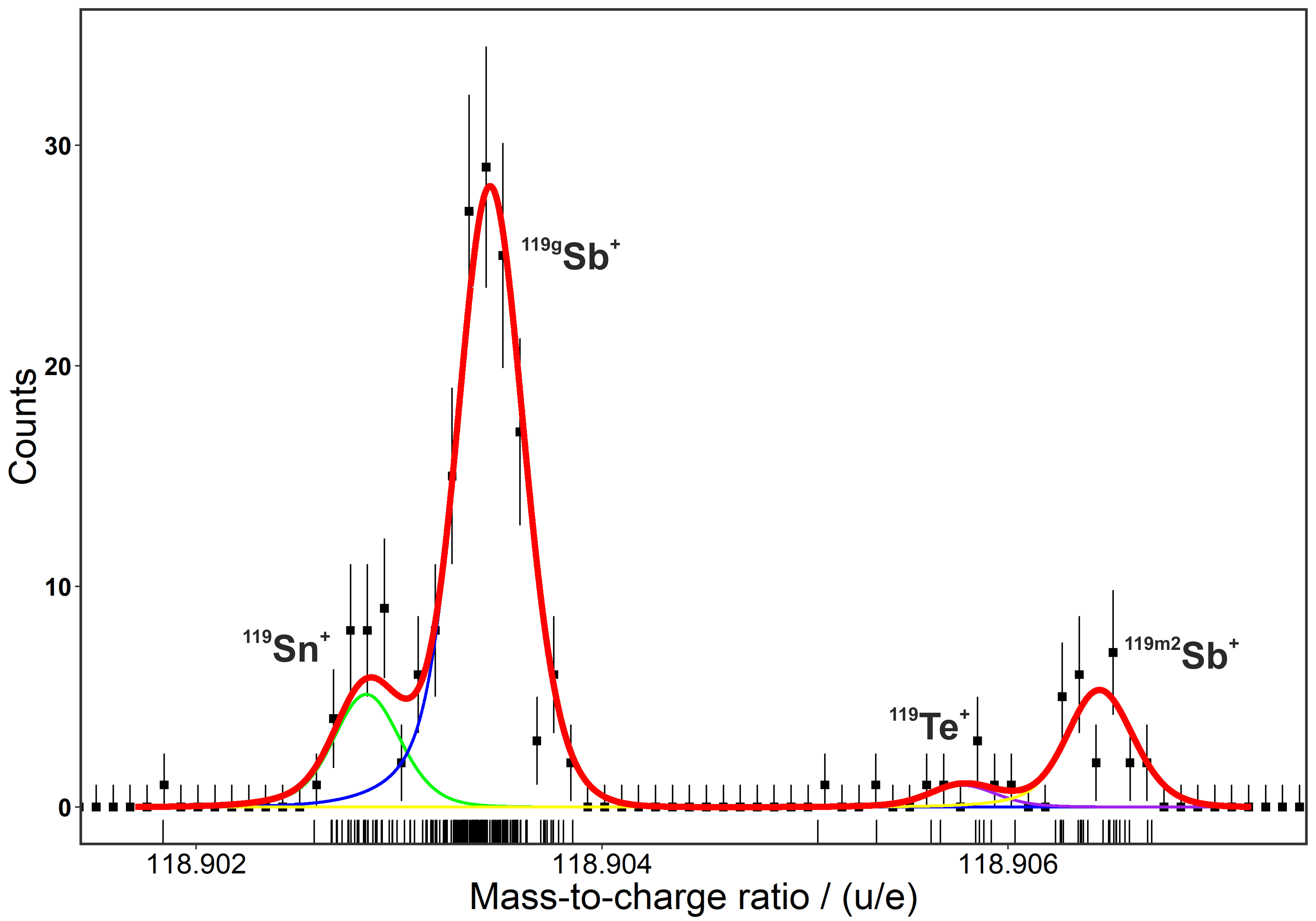}
}
\caption{Mass spectrum of $^{119m2}$Sb, $^{119g}$Sb, $^{119}$Te and $^{119}$Sn measured with a mass resolving power of 290,000. The red solid line represents the overall fit, whereas the green, blue, yellow and pink lines represent fits to the separate isotope peaks, enabling extraction of isotope masses and counts. The $^{119}$Te and $^{119}$Sn long-living isomeric states are indistinguishable from the respective ground states. The un-binned data is shown as well.}
\label{fig:2}       
\end{figure}

In Fig.~\ref{fig:2} we show a measured mass spectrum of $^{119m2}$Sb, $^{119g}$Sb, $^{119}$Te and $^{119}$Sn, resulting from the fragment beam described above. Note that $^{119}$Te and $^{119}$Sn both have long-lived isomeric states (260.96 keV, 4.70 days and 89.531 keV, 293.1 days, respectively \cite{Audi2016}). Based on \cite{Ayet2019}, the excitation energy of the $^{119}$Sn isomeric state is below the FRS-IC resolving power, and although that of $^{119}$Te is in principle resolvable, it was not possible to distinguish it with the low statistics and primary beam conditions of this work. Therefore, the mass spectra are fitted to a single peak, and the results are compared to the ground state literature values. 

As in Fig.~\ref{fig:1}, the fit curves are from the procedure described in \cite{Ayet2019}. One can observe the separate peak fits for the distinguishable overlapping peaks ($^{119g}$Sb and $^{119}$Sn), which give the mass values and the number of counts for each isotope. 
The mass of $^{119g}$Sb and the excitation energy of $^{119m2}$Sb are measured directly for the first time. Our results and several measurement details are given in Table~\ref{tab:FRSIC_Masses}. The measured masses of $^{119g}$Sb, $^{119}$Sn and $^{119}$Te are consistent with the literature values. In Section~\ref{sec:3.2} we compare our $^{119m2}$Sb excitation energy result to literature and discuss its impact.

\subsection{Insights from $^{119m2}$Sb mass measurement}
\label{sec:3.2}

The second isomeric level of $^{119}$Sb, namely $^{119m2}$Sb ($t_{1/2}=850\pm90$ ms), was first identified in 1979 \cite{Shroy1979}.
Its excitation energy and spin assignment could not be firmly established so far in spite of multiple attempts \cite{Shroy1979,Lunardi1987,Porquet2005,Moon2011}. This was because of
clear methodical problems due to correlations and/or multiple $\gamma$ rays present in the spectra at almost the same energy, as well as the existence of another isomeric state at 2553.6$\pm0.3$ keV. 
The adopted values for $^{119m2}$Sb in the Atlas of Nuclear Isomers \cite{Jain2015} and ENSDF \cite{Symochko2009} are an excitation energy of 2841.7$\pm0.4$ + x keV, where x is to be determined, and spin and parity of 27/2$^+$. The adopted assignment for the 2841.7 keV level is $I^{\pi} = 21/2^+$ \cite{Symochko2009}. 

Our result for the $^{119m2}$Sb isomeric state excitation energy,
2799$\pm30$ keV (Table~\ref{tab:FRSIC_Masses}), is the first experimental data that is based on a direct mass measurement, and is 1.4 $\sigma$
below 2841.7$\pm0.4$ keV. It suggests that the above-mentioned
parameter x is consistent with 0, and the 2841.7 keV level
itself is the long-lived isomeric state, as was presented originally in Ref. \cite{Lunardi1987} (fusion reaction experiment) and adopted in Ref. \cite{Porquet2005} (induced fission experiment). 

The experiments of Refs. \cite{Lunardi1987,Porquet2005} were based on $\gamma$ spectroscopy, and both designated $I^{\pi} = 25/2^+$ to the 2841.7 keV level, due to the conversion coefficient value of the 288.2 keV transition depopulating the isomer.

We thus conclude that our result confirms the $^{119m2}$Sb isomeric state designation of Refs. \cite{Lunardi1987,Porquet2005}, in contradiction to the adapted assignment \cite{Jain2015,Symochko2009}.

\begin{table*}[t]
\centering
    \begin{tabular}{ccccccc} \hline\hline
    Nuclide & Half-life & ME$_\mathrm{FRS-IC}$ & ME$_\mathrm{AME16}$ & ME$_\mathrm{FRS-IC}$ - ME$_\mathrm{AME16}$ & Number \\
     & & E$_\mathrm{exc, FRS-IC}$ & E$_\mathrm{exc}$ \cite{Jain2015} & E$_\mathrm{exc, FRS-IC}$ - E$_\mathrm{exc}$ \cite{Jain2015} & of events \\ 
     & & / keV & / keV & / keV &  \\ \hline
		$^{119g}$Sb & $38.19\pm0.22\>$h & $-89482\pm17$ & -89474 $\pm$ 8 & -8 $\pm$ 19 & 857 \\
        \multirow{2}{*}{$^{119m2}$Sb} & 850 $\pm$ 90$\>$ms & \multirow{2}{*}{2799 $\pm$ 30} & \multirow{2}{*}{$2841.7\pm0.4$ + x} & \multirow{2}{*}{-43 $\pm$ 30} & \multirow{2}{*}{467} \\
       & 776 $\pm$ 181$\>$ms$^{*}$&  &  &  &  &  \\
        $^{119}$Sn & Stable &  -90100 $\pm$ 36 & -90065.0 $\pm$ 0.7 & -35 $\pm$ 36 & 237 $^{**}$\\
        $^{119}$Te & 16.05 $\pm$ 0.05$\>$h & -87310 $\pm$ 177 & -87181 $\pm$ 8 & -129 $\pm$ 177 &  25$^{**}$ \\ \hline\hline    \end{tabular}
\caption[Direct Mass Measurements at the FRS-IC]{Results of direct mass measurements performed at the FRS-IC. The uncertainties shown correspond to the total experimental uncertainty. Literature values are from \cite{Audi2016,Jain2015,Symochko2009,Wang2017}. In the MR-TOF-MS, the total time-of-flight was 18.9 ms, and the number of turns was 560. The reference ion in all these measurements was $^{12}$C$_{2} ~^{19}$F$_{5}$ (A=119). $^{*}$ Literature and measured value given. $^{**}$ Total number of events for the unresolved ground and isomeric states. The literature values of $^{119}$Sn and $^{119}$Te are for their ground states.}
\label{tab:FRSIC_Masses}
\end{table*}
\renewcommand{\arraystretch}{1}

\subsection{Long-term containment in the CSC}
\label{sec:3.3}
We investigated long-term containment in the CSC with the stable and relatively long-living isotopes of the $^{228}$Th decay chain from the CSC internal radioactive $\alpha$-recoil source. Our goal was to demonstrate isotope containment in the CSC without non-nuclear-decay losses (e.g., charge exchange, molecular formation).

\begin{figure}
\resizebox{0.48\textwidth}{!}{%
 \includegraphics{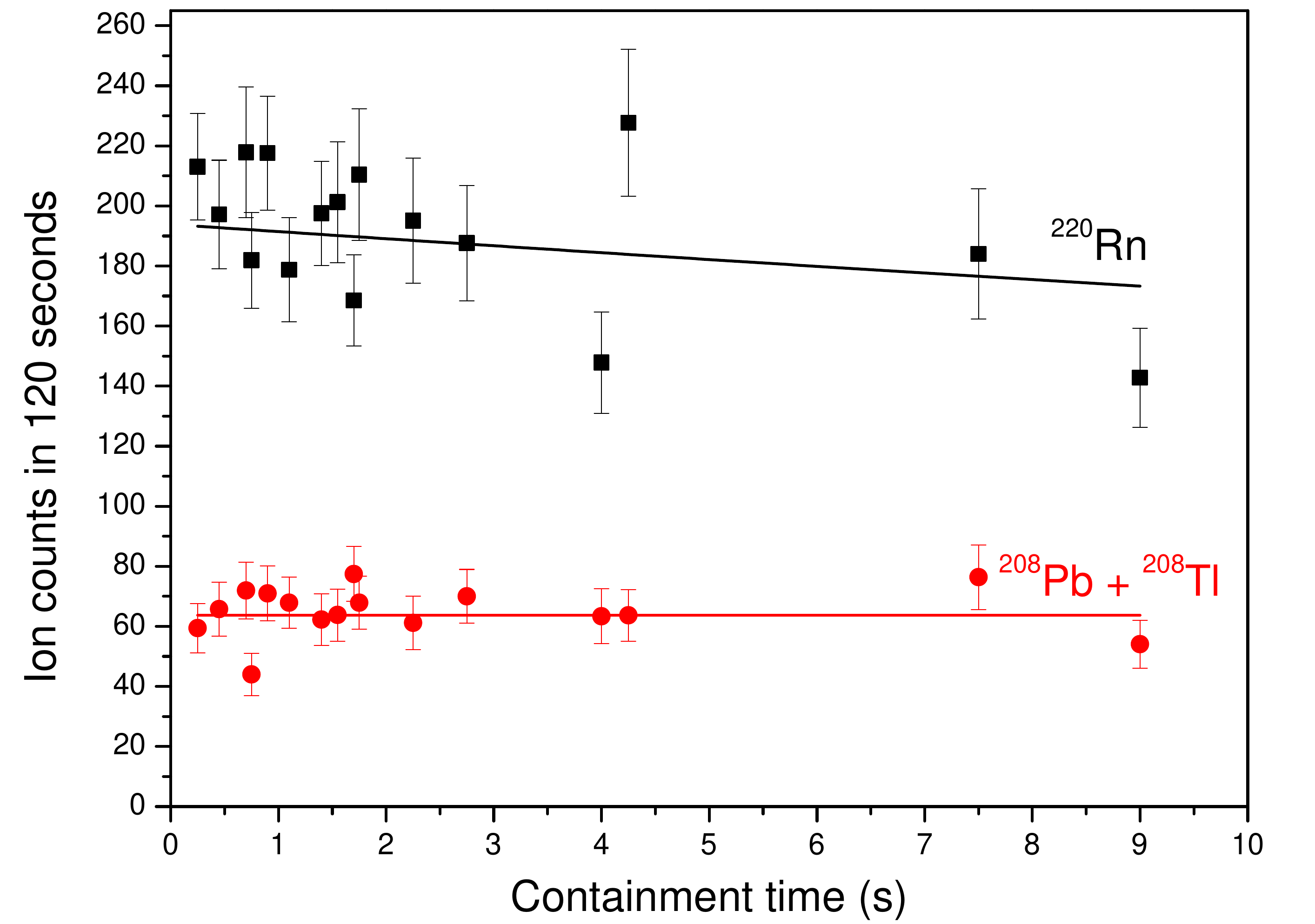}
}
\caption{$^{220}$Rn and the sum of $^{208}$Pb and $^{208}$Tl counts as a function of CSC containment time. Solid lines are fits with the literature time constants. The $^{208}$Pb and $^{208}$Tl counts are consistent with the expected stable behavior and the $^{220}$Rn counts are consistent with the literature value t$_{1/2}$ = 55.6 seconds.} 
\label{fig:3}       
\end{figure}

We measured the amount of a combination of $^{208}$Pb (stable) and $^{208}$Tl (t$_{1/2}$ = 3.053 minutes) as a function of CSC containment time, which we varied from 0.2 seconds up to 9 seconds. The MR-TOF-MS was set to low mass resolving power, so these two isotopes were not resolved. As can be seen from the red data points in Fig.~\ref{fig:3}, which are fitted to a constant (red solid line), there are no non-nuclear-decay losses for Pb and Tl ions over these long containment times.

$^{220}$Rn (t$_{1/2}$ = 55.6 seconds) ions are also detected in the same measurement. As can be seen from the black data points in Fig.~\ref{fig:3}, which are fitted to an exponential decay curve with the literature half-life (black solid line), these results are also consistent with negligible non-nuclear losses.

We have thus shown that the CSC can contain isotopes up to about 10 seconds. This indicates the CSC is extraordinarily clean, with relevant impurities on a level lower than 10$^{-10}$.
Note that we limited our measurements to this containment time due to practical constraints, so the actual upper limit may be higher.
\subsection{Half-life and branching ratio measurement}
\label{sec:3.5}

We demonstrated simultaneous precursor decay and recoil growth via the decay of $^{216}$Po ((t$_{1/2}$ = $145\pm2$ ms) to $^{212}$Pb (t$_{1/2}$ = 10.64 hours), both part of the decay chain of the internal $^{228}$Th radioactive recoil source in the CSC.

During data analysis, we realized that in the experiment there was an error in the CSC RF carpet settings, which resulted in unusually long transport of the isotopes along the RF carpet, $\sim$170 ms instead of a few ms. 
This meant that in these measurements, all isotopes were contained within $\sim$ 0.1 mm from the RF carpet for an unknown duration ranging between $0$ and \mbox{$\sim$170 ms} (namely, $\sim$85$\pm$85 ms), after the 70 ms drift from the $^{228}$Th source situated at the upstream edge of the CSC. We were able to correct the data for that error, albeit inducing a large systematic uncertainty on the number of $^{216}$Po counts, especially for containment times below \mbox{$\sim$200 ms}. A dead-time correction was also applied to the data, further increasing the systematic uncertainty.

In the error correction procedure we used the literature half-life of $^{216}$Po, in order to account for possible decays during the containment of the isotopes near the RF carpet. This was done only for the counts with containment times below \mbox{$\sim$200 ms}. The effect of including the literature half-life on the final half-life result was small with respect to the main uncertainty component, which is the $\pm$85 ms uncertainty in the transport time of the isotopes near the RF carpet.

We plotted the precursor and recoil counts normalized to their sum and performed a global fit on all data points to Eqs.~\ref{eq:counts_precursor} and ~\ref{eq:counts_recoil}, according to the procedure described in Section~\ref{sec:2.3}. The relatively long half-life of $^{212}$Pb justified using Eq. ~\ref{eq:counts_recoil} instead of Eq.~\ref{eq:counts_recoil_decay}.
No a-priori assumptions are taken, since we include in the normalization all isotopes whose decay from the investigated precursor have a positive Q-value. For $^{216}$Po, only $\alpha$ decay to $^{212}$Pb meets this criterion.

\begin{figure}
\resizebox{0.48\textwidth}{!}{%
  \includegraphics{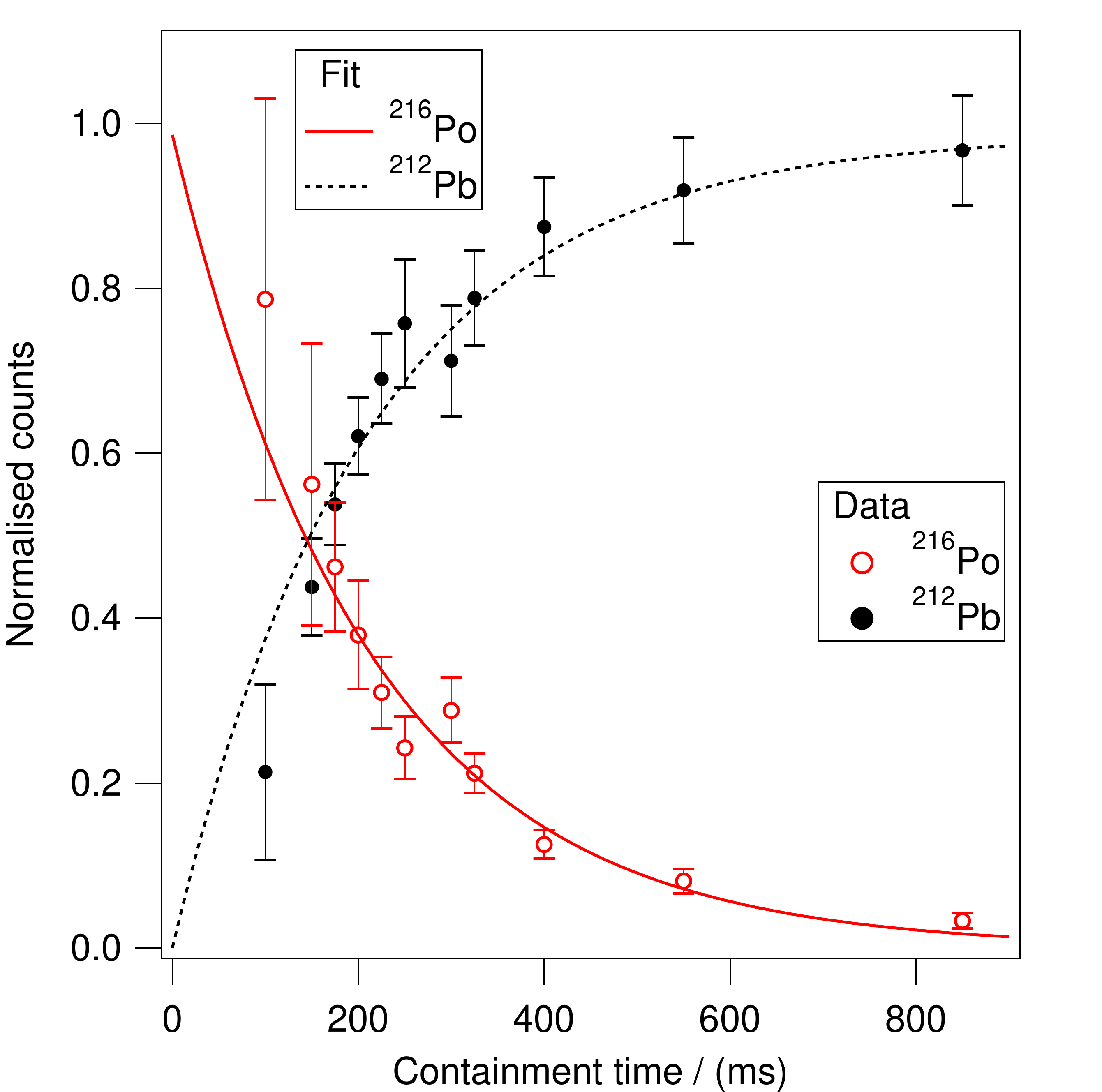}
}
\caption{$^{216}$Po and $^{212}$Pb counts, normalized to their sum, as a function of CSC containment time, up to about 1 second. The curves represent a global fit to both data sets to Eqs.~\ref{eq:counts_precursor} and ~\ref{eq:counts_recoil}.} 
\label{fig:5}       
\end{figure}

In Fig.~\ref{fig:5} we depict $\alpha$ decay of $^{216}$Po to $^{212}$Pb in the self-normalizing presentation described above. 
The effect of the CSC setup error discussed above is visible in the increased uncertainties of both the $^{216}$Po and $^{212}$Pb counts, especially at containment times below \mbox{$\sim$200 ms}, as expected. The $^{212}$Pb counts were further corrected for the 50\% loss at the RF carpet vicinity due to their long range (see Section~\ref{sec:2.4}).

It is evident by eye that the two trends are consistent with a precursor decay and its recoil growth. The results of the global fit in Fig.~\ref{fig:5} are given in Table~\ref{tab:216Po_Fit}. Per the defined constraint, the sum of $y_{0r1}$ and $A_p$ is consistent with unity. A precursor half-life of $145\pm11$ ms is obtained, consistent with the literature value ($145\pm2$ ms \cite{Audi2016}).

To test the effect of the global fit, we also fitted the $^{216}$Po curve separately to Eq.~\ref{eq:counts_precursor}, without any constraint with respect to the $^{212}$Pb data. The half-life result in this case is $158\pm15$ ms, consistent with the global fit results, but with a larger uncertainty. The advantage of including the recoil data is evident.

In addition, we note that we fitted the un-normalized $^{216}$Po data to Eq.~\ref{eq:counts_precursor}, and also obtained a half-life result that is consistent with the literature value, and with an uncertainty that is similar to the result of the independent fit to the normalized data.

\begin{table}[t]
\centering
    \begin{tabular}{ccc} \hline\hline
     \multirow{2}{*}{$^{216}$Po} & $A_P$ & 1.00 $\pm$ 0.10 \\
        & $t_{1/2}$ [ms] & 145 $\pm$ 11 \\
    \hline
    \multirow{1}{*}{$^{212}$Pb} & $y_{0r1}$ & 0.00 $\pm$ 0.10  
     \\ \hline\hline
    \end{tabular}
\caption[216Po Fit]{$^{216}$Po and $^{212}$Pb global fit results}
\label{tab:216Po_Fit}
\end{table}

We further used our method to measure the half-life and decay branching ratios of $^{119m2}$Sb, obtained from the fragment beam.
For $^{119m2}$Sb it is energetically possible to decay to $^{119}$Sn ($\beta^{+}$ decay) and $^{119}$Te ($\beta^{-}$ decay), in addition to the $\gamma$-decay to $^{119g}$Sb. We plotted the normalized counts of the precursor $^{119m2}$Sb, and its energetically possible decay products. Such a presentation is independent of absolute normalization of the incoming fragment beam from the FRS. Because of the limited primary beam energy, we obtained data with limited statistics and significant beam-induced contamination, including $^{119}$Sn and $^{119}$Te. 

\begin{figure}
\resizebox{0.48\textwidth}{!}{%
 \includegraphics{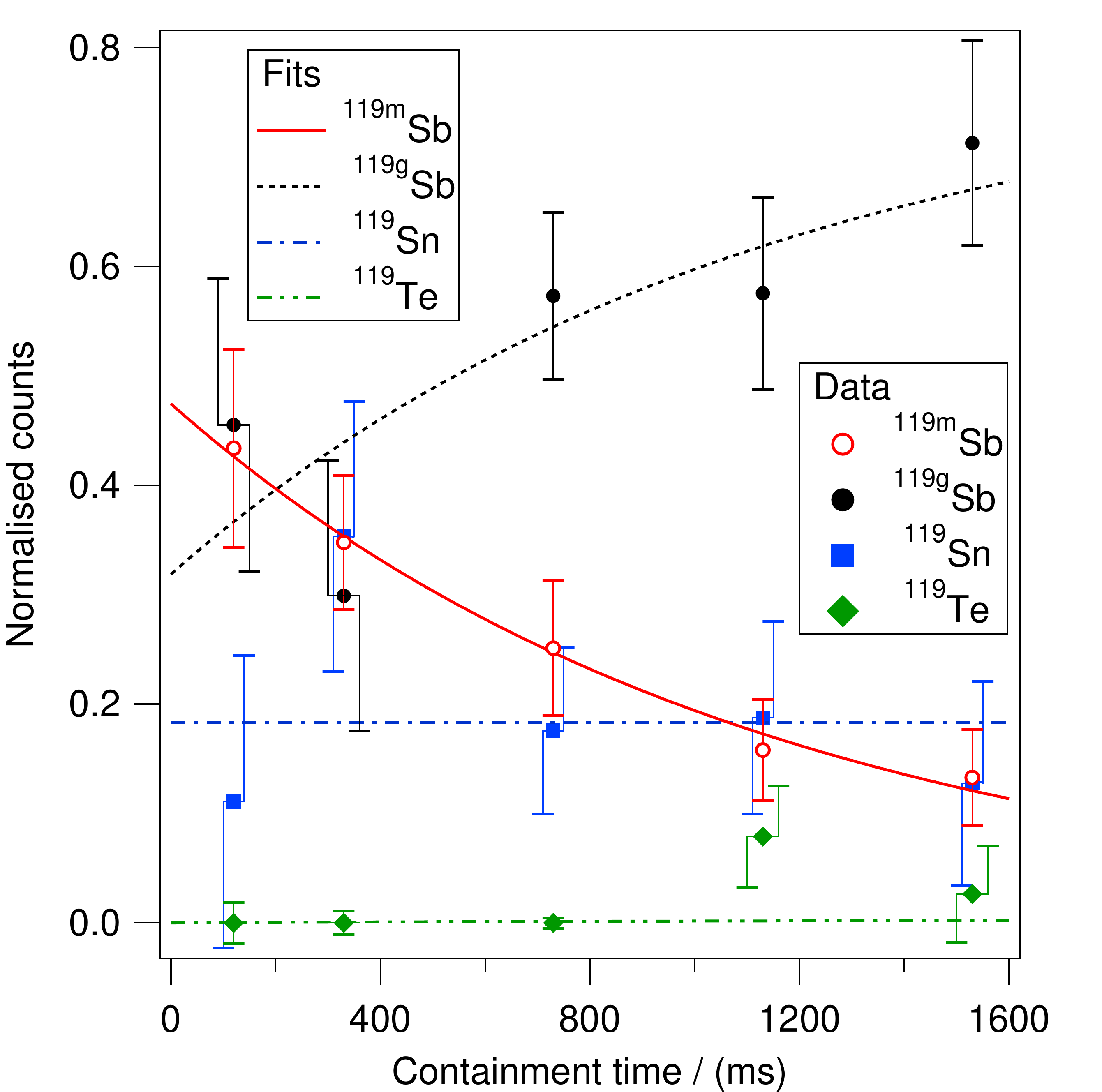}
}
\caption{Normalized counts of the precursor $^{119m2}$Sb and its energetically possible decay recoils $^{119g}$Sb ($\gamma$-decay), $^{119}$Sn ($\beta^{+}$ decay) and $^{119}$Te ($\beta^{-}$ decay), as a function of containment time, up to about 1.5 seconds. Counts of precursors and recoil candidates are normalized to their total sum. Lines represent a global fit of all data sets to Eqs.~\ref{eq:counts_precursor} and ~\ref{eq:counts_recoil}.} 
\label{fig:6}       
\end{figure}

The results are shown in Fig.~\ref{fig:6}. The counts are normalized to the sum of the precursor and all decay recoil candidates. The total amount of events contributing to Fig.~\ref{fig:6} is approximately half of the total amount in Table~\ref{tab:FRSIC_Masses}, because not all events were recorded under the same conditions. The whole data set is fitted with a global fit to Eqs.~\ref{eq:counts_precursor} and ~\ref{eq:counts_recoil}, according to the procedure described in Section~\ref{sec:2.3}. The half-lives of all possible recoils are long enough to justify not using Eq.~\ref{eq:counts_recoil_decay}. 
The fit results are given in Table~\ref{tab:119Sb_Fit}.

The $^{119m2}$Sb half-life given in Table~\ref{tab:119Sb_Fit} is consistent with the literature value ($850\pm90$ ms \cite{Jain2015}). It is clear that only $^{119g}$Sb behaves as a recoil isotope, growing according to Eq.~\ref{eq:counts_recoil} with its precursor's half-life (implying branching ratio of unity for $\gamma$ decay), whereas the other two isotopes exhibit a constant temporal behavior, consistent with a branching ratio of zero for $\beta^+$ and $\beta^-$ decays.

Our data is thus the first experimental proof that $^{119m2}$Sb decays only via $\gamma$-ray emission. Unlike $\gamma$-spectroscopy experiments that did not observe transitions that are indicative of $\beta^+$ or $\beta^-$ decays, we showed positively that there is no growth of the $\beta^+$ nor $\beta^-$ decay-recoils.

We can further extract from Table~\ref{tab:119Sb_Fit} the $^{119}$Sb isomer ratio ($IR$) in the FRS fragment beam, by $IR=\frac{y_{0r1}}{A_P}$. The result we obtain is $0.7\pm0.3$. Note that this result is specific to the primary beam conditions of this experiment, described above.

The global fit with all possible recoil candidates is not optimal for the extraction of the half-life of $^{119m2}$Sb, since it includes input from isotopes that in retrospect have no physical connection to this precursor. It thus merely contributes noise to the fit and increases the uncertainty of the fitted parameters. Therefore, we determine generally that for the extraction of the precursor half-life with the smallest uncertainty, one should repeat the procedure with only the recoils that have a finite branching ratio. We thus re-fitted only the $^{119m2}$Sb and $^{119g}$Sb curves, and obtained for $^{119m2}$Sb a half-life value of $776\pm181$ ms (see Table~\ref{tab:FRSIC_Masses}). This result is consistent with the value in Table~\ref{tab:119Sb_Fit}, and is indeed with a smaller uncertainty.

\begin{table}[t]
\centering
    \begin{tabular}{ccc} \hline\hline
    \multirow{2}{*}{$^{119m2}$Sb} & $A_P$ & 0.48 $\pm$ 0.08 \\
        & $t_{1/2}$ [ms] & 784 $\pm$ 203 \\
    \hline
    \multirow{2}{*}{$^{119g}$Sb} & $y_{0r1}$ & 0.33 $\pm$ 0.12 \\
        & $n_{r1}$ & 0.99 $\pm$ 0.51 \\
    \hline
    \multirow{2}{*}{$^{119}$Sn} & $y_{0r2}$ & 0.19 $\pm$ 0.12 \\
        & $n_{r2}$ & 0.00 $\pm$ 0.45 \\
    \hline
    \multirow{2}{*}{$^{119}$Te} & $y_{0r3}$ & 0.00 $\pm$ 0.02 \\
        & $n_{r3}$ & 0.01 $\pm$ 0.08 
     \\ \hline\hline
    \end{tabular}
\caption[119Sb Fit]{Global fit results for $^{119m2}$Sb decay and all its possible recoils}
\label{tab:119Sb_Fit}
\end{table}

\section{Further developments and future measurements}
\label{sec:4}
The method presented in this article is well-suited for the measurement of decay branching ratios of nuclei, which decay by more than one channel.

However, as discussed in Section~\ref{sec:2.4}, for the method to be operational for all relevant nuclear decay types, we must ensure that the contained isotopes of interest spend the vast majority of their time in the CSC bulk, far away (several cm) from any of the CSC electrodes.
We describe in this subsection an improvement that will ensure that decays takeplace mostly within the CSC bulk, thus minimizing recoil losses on electrodes.

Further in this Section, we discuss possible future measurements of half-lives and decay branching ratios with our novel method.

\subsection{Dynamic storage of ions in the CSC bulk}
\label{sec:4.1}
We have devised a new 3-step scheme for isotope dynamic containment in the CSC bulk, in order to ensure minimal or practically no recoil loss due to collision with the CSC inner walls.
In step 1, the ions from the beam (short pulse, $\sim$ms) are stopped in the CSC, transported via the DC field to the RF carpet within twice the mean extraction time ($t \sim 2\times t_{extr} \sim 50$ ms \cite{Plass2013}), and bunched near the extraction nozzle, which is on high potential to block extraction. In step 2, the DC field is switched to a low negative value with respect to the nozzle, and within ~10 ms the isotopes are ~1 cm away from the nozzle. The nuclides diffuse and drift in the weakening electric field towards the CSC center, for a duration controllable from milliseconds to seconds. In step 3, the DC field is switched back to its nominal value and the nozzle potential is lowered, extracting the isotopes from the CSC within a few milliseconds. 

\begin{figure*}[ht]
\resizebox{1\textwidth}{!}{%
  \includegraphics{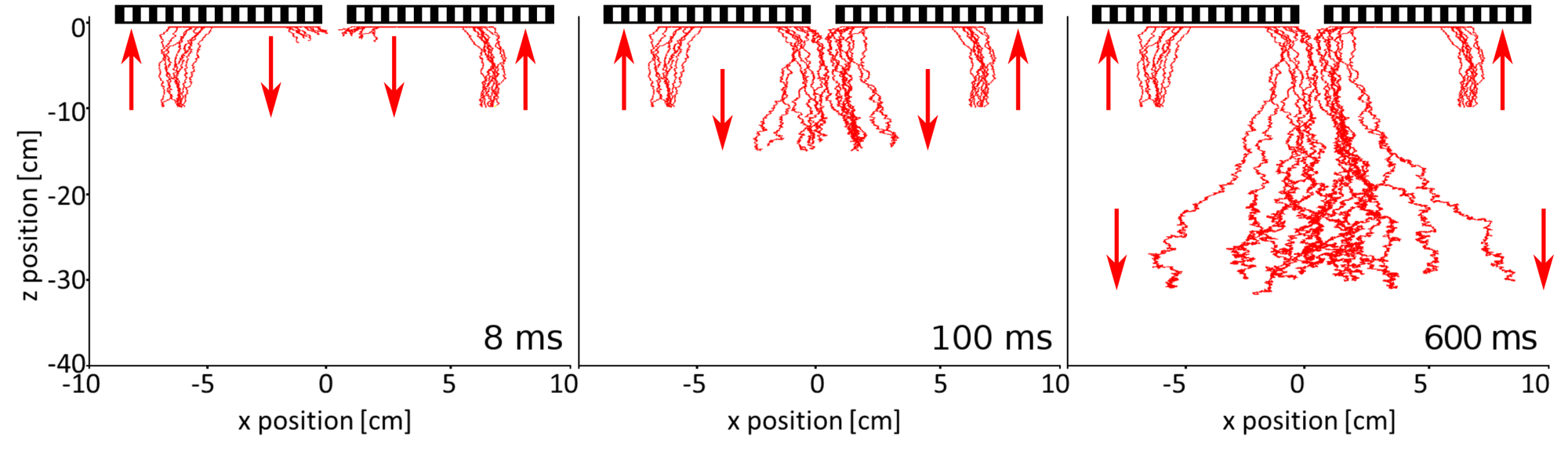}
}
\caption{ITSIM simulation results for dynamic storage in the CSC. Each panel depicts the CSC downstream section. The RF carpet is marked as a hash. The nozzle diameter is not to scale. The vertical axis (z) is from -40 to 0 cm (top), and the horizontal axis (x) is from -10 to +10 cm. The nozzle is at center-top, z = -0.5 cm. The CSC is 105 cm long (z) and 25 cm in diameter (x). Left: from a late part of Step 1, to 8 ms into Step 2, when the isotopes are $\sim$1 cm away from the nozzle. Middle: Time extended to 100 ms into Step 2. Right: Isotope trajectories from 100 to 600 ms into Step 2 (bulk containment). Incoming and outgoing trajectories are x-separated for clarity. Arrows indicate isotopes general direction.}
\label{fig:7}       
\end{figure*}

Snapshots from a simulation via ITSIM \cite{Wu2006,Plass2008-1} of the isotope trajectories during the above process are given in Fig.~\ref{fig:7}.
The simulation conditions were: He gas at 100~$\mu$g/cm$^3$, 155 mbar, and 80 K, RF carpet voltage of 120 V peak-to-peak at 6 MHz, with a radial DC gradient of 4 V/cm, DC longitudinal field of 10 V/cm in step 1 and -0.5 V/cm in step 2. Ions with mass-to-charge ratio of 219 u/e were used, with a macroscopic collision model, which implies that ion motion through the gas is governed by their mobility value, set to 17.5 cm$^2$/(V$\cdot$s). 

The simulation shows that the isotopes are kept at a safe distance from the CSC inner walls for most of their duty cycle. If longer containment times are needed, Step 1 and 2 may be cycled. It is important to emphasize that this scheme is possible only because of the small diffusion of the isotopes in the CSC, and due to its high pressure and low temperature, which enables a long containment time. 

\subsection{Half-lives and decay branching ratio measurements}
\label{sec:4.2}
The novel measurement method presented here is a powerful mean to investigate masses, half-lives and decay branching ratios of numerous exotic nuclides, in ground and isomeric states, with half-lives ranging from a few tens of milliseconds to a few seconds. In the future CSC for the Low-Energy Branch of the Super-FRS \cite{Geissel2003}, shorter-lived isotopes will be reachable, since its extraction time will be improved to $\sim$5 ms \cite{Dickel2016}.

Furthermore, branching ratios to spontaneous and $\beta$-delayed fission could be investigated in the future CSC. This is because its density will be as high as 200~$\mu$g/cm$^3$ \cite{Dickel2016}, enabling almost full fission fragment stopping and containment (the $\sim$100 MeV fission fragments' range will be $\sim$12 cm ). This will enable unique research opportunities in the field of fission, because our MR-TOF-MS will provide isotope yield distributions and isomer yield ratios, not only mass distributions.

As our method relies on nuclide identification by direct mass measurements and not by gamma de-excitation lines, it may be the method of choice for nuclides far away from stability, whose excited energy levels are not well-known, or not known at all. It can also provide independent and complementary data for exotic nuclides whose properties were measured via gamma de-excitation, but their values (or their physical interpretation) is under debate, e.g. the excitation energy of the isomeric state $^{119m2}$Sb discussed in this work. 

According to the Atlas of Nuclear Isomers \cite{Jain2015}, there are at least 20 more isomeric states that could be investigated with our method at the FRS-IC at GSI (ranging from Co to Ra isotopes, with half-lives in the above-mentioned region and resolvable excitation energies), whose excitation energy and quantum numbers are still ambiguous. Specific examples include $^{178}$Ta with two such isomeric levels at 1467.82+x keV (15$^-$, 58 ms) and 2902.2+x keV (21$^-$, 290 ms), and $^{188}$Bi and $^{199}$Rn for whom also the branching ratio between $\alpha$ and $\beta^+$ decays is yet unmeasured. 

A particularly interesting isomeric state for investigation with our method is the $6490\pm500$ keV isomeric state of $^{94}$Ag (21$^+$, 400 ms), which attracted substantial attention as the so-far only nuclear state to decay by both one-proton and two-proton direct emission \cite{Mukha2006}, in addition to regular $\beta^+$ decay, and $\beta$-delayed one- and two-proton emission. This rich variety of decay channels, and the fact that its direct two-proton direct emission channel is still under scrutiny \cite{Audi2016}, makes this nuclide especially attractive for investigation with a new independent method. An experiment that will investigate N=Z nuclei (including $^{94}$Ag) in the FRS Ion Cather is foreseen in the near future \cite{Plass2018}.

$\beta$-delayed nucleon emission may be a field where our method could make a significant impact. For very neutron-deficient nuclides, the direct identification of the recoils can serve to differentiate between direct and $\beta$-delayed proton emission, especially for cases where the precursor and recoil candidates are not sufficiently well known to rule out one of the channels due to negative Q-value.

$\beta$-delayed single- and multi-neutron emission is of particular interest to nuclear structure and r-process nucleosynthesis, and is more challenging experimentally than $\beta$-delayed proton emission due to the relative complexity of quantitative neutron detection. There are thus numerous ongoing and near-future campaigns to measure $\beta$-delayed single- and multi-neutron emission branching ratios 
\cite{Dillman2018,Yee2013,Evdokimov2012,Miyatake2018,BDN-IAEA-CRP}, including a planned experiment utilizing our method at GSI \cite{Mardor2018}.

This new paradigm, of measuring nuclear processes by mass measurements of the outgoing nuclides in an Ion Catcher, may be expanded from decays to reactions. Identifying and counting the outgoing nuclides can be used to infer simultaneously total reaction cross sections of all existing channels for particular projectile-target combinations. A first experiment in this direction is planned for investigating multi-nucleon transfer reactions of $^{238}$U on several medium- and heavy-ion targets that will be installed in the CSC of the FRS-IC \cite{Dickel2018}.

\section{Conclusions}
\label{sec:5}
We described a new method to measure simultaneously masses, isomer excitation energies, Q-values, half-lives and branching ratios, which was developed and demonstrated in the FRS-IC at GSI. The high fragment energy in the FRS makes it the only facility where this method can be implemented. The method includes a first use of a stopping cell as an ion trapping device, for controllable duration from a few tens of ms up to $\sim$10 seconds.

We demonstrated the method's feasibility with isotopes from an internal $\alpha$-recoil radioactive source in the stopping cell, and from the FRS fragment beam. For both cases, the expected temporal behavior of precursors and recoils was shown, and half-lives and branching ratios that are consistent with the literature were obtained.

The method requires only very general prior knowledge of the investigated isotopes. This was shown nicely in this work, where while demonstrating the method, we performed the first direct mass measurement of the ground state and second isomeric state of $^{119}$Sb, whose excitation energy and quantum numbers are still not firmly established. Our result for the $^{119m2}$Sb excitation energy suggests that the isomeric state is the energy level 2841.7 keV, with $I^{\pi}=25/2{^+}$ (based on previous $\gamma$-ray spectroscopy experiments \cite{Lunardi1987,Porquet2005}), in contradiction to the adopted assignment \cite{Jain2015,Symochko2009}. We also confirmed positively that $\gamma$ decay is the only branch of the depopulation of this isomeric state.

We have developed a dynamic storage procedure that will ensure isotope containment mainly in the CSC bulk, enabling research of all possible decay channels, including spontaneous and $\beta$-delayed fission. 

This method opens the door for a vast range of decay and reaction processes, in an independent and complementary way with respect to existing measurement methods.
%
%
%
%
\section{Authors contributions}
All the authors were involved in the preparation of the manuscript.
All the authors have read and approved the final manuscript.
\section{Acknowledgements}
This work was supported by the German Federal Ministry for Education and Research (BMBF) under under contracts no.\ 05P12RGFN8 and 05P15RGFN1, by Justus-Liebig-Universit{\"a}t Gie{\ss}en and GSI under the JLU-GSI strategic Helmholtzpartnership agreement, by HGS-HIRe, and by the Hessian Ministry for Science and Art (HMWK) through the LOEWE Center HICforFAIR. I. Mardor acknowledges support from the Israel Ministry of Energy, Research Grant No. 217-11-023.
%
%
%
%

\bibliographystyle{epj}       
\bibliography{References} 


\end{document}